\documentclass[sigconf]{acmart}
\usepackage{booktabs} % For formal tables
\usepackage{amsmath}
\usepackage{amsfonts}
\usepackage{comment}
\usepackage{footmisc}
\usepackage{mathrsfs}
\usepackage{amssymb}
\usepackage{graphicx}
\usepackage{subfigure}
\usepackage{multirow}
\usepackage{algorithm}
\usepackage{algorithmic}
\usepackage{multicol}  
\usepackage{enumitem}
\usepackage{array}
\usepackage{float}
\usepackage{hyperref}
\AtBeginDocument{%
	\providecommand\BibTeX{{%
			\normalfont B\kern-0.5em{\scshape i\kern-0.25em b}\kern-0.8em\TeX}}}

\copyrightyear{2020}
\acmYear{2020}
\setcopyright{acmcopyright}
\acmConference[KDD '20]{Proceedings of the 26th ACM SIGKDD Conference on Knowledge Discovery and Data Mining}{August 23--27, 2020}{Virtual Event, CA, USA}
\acmBooktitle{Proceedings of the 26th ACM SIGKDD Conference on Knowledge Discovery and Data Mining (KDD '20), August 23--27, 2020, Virtual Event, CA, USA} 
\acmPrice{15.00}
\acmDOI{10.1145/3394486.3403384}
\acmISBN{978-1-4503-7998-4/20/08}

\begin{document}
\fancyhead{}
\title{Jointly Learning to Recommend and Advertise}
\author{Xiangyu Zhao}
\affiliation{
	\institution{Michigan State University}
}
\email{zhaoxi35@msu.edu}

\author{Xudong Zheng}
\affiliation{
	\institution{Bytedance}
}
\email{zhengxudong.alpha@bytedance.com}

\author{Xiwang Yang}
\affiliation{
	\institution{Bytedance}
}
\email{yangxiwang@bytedance.com}

\author{Xiaobing Liu}
\affiliation{
	\institution{Bytedance}
}
\email{will.liu@bytedance.com}

\author{Jiliang Tang}
\affiliation{
	\institution{Michigan State University}
}
\email{tangjili@msu.edu}

\renewcommand{\shortauthors}{Xiangyu Zhao, et al.}

\begin{abstract}
	Online recommendation and advertising are two major income channels for online recommendation platforms (e.g. e-commerce and news feed site). However, most platforms optimize recommending and advertising strategies by different teams separately via different techniques, which may lead to suboptimal overall performances. To this end, in this paper, we propose a novel two-level reinforcement learning framework to jointly optimize the recommending and advertising strategies, where the first level generates a list of recommendations to optimize user experience in the long run; then the second level inserts ads into the recommendation list that can balance the immediate advertising revenue from advertisers and the negative influence of ads on long-term user experience. To be specific, the first level tackles high combinatorial action space problem that selects a subset items from the large item space; while the second level determines three internally related tasks, i.e., (i) whether to insert an ad, and if yes, (ii) the optimal ad and (iii) the optimal location to insert. The experimental results based on real-world data demonstrate the effectiveness of the proposed framework. We have released the implementation code to ease reproductivity. 
\end{abstract}


\maketitle

\section{Introduction}
\label{sec:introduction}
Practical e-commerce or news-feed platforms generally expose a hybrid list of recommended and advertised items (e.g. products, services, or information) to users, where recommending and advertising algorithms are typically optimized by different metrics~\cite{feng2018learning}. The recommender systems (RS) capture users' implicit preferences from historical behaviors (e.g. clicks, rating and review) and generate a set of items that best match users' preferences. Thus, RS aims at optimizing the user experience or engagement.  While advertising systems (AS) assign the right ad to the right user on the right ad slots to maximize the revenue, click-through rate (CTR) or return on investment (ROI) from advertisers. Thus, optimizing recommending and advertising algorithms independently may lead to suboptimal overall performance since exposing more ads to increase advertising revenue has a negative influence on user experience, vice versa. Therefore, there is an increasing demand for developing a uniform framework that jointly optimizes recommending and advertising, so as to optimize the overall performance~\cite{zhao2019ads}. 

Efforts have been made on displaying recommended and advertised items together.  They consider ads as recommendations, and rank all items in a hybrid list to optimize the overall ranking score~\cite{wang2018learning}. However, this approach has two major drawbacks. First, solely maximizing the overall ranking score may result in the suboptimal advertising revenue. Second, in the real-time bidding (RTB) environment, the vickrey-clarke-groves (VCG) mechanism is necessary to calculate the bid of each ad in the list, which suffers from many serious practical problems~\cite{rothkopf2007thirteen}. Therefore, it calls for methods where we can optimize not only the metrics for RS and AS separately, but also the overall performance. Moreover, more practical mechanisms such as generalized-second-price (GSP) should be considered to compute the bid of each ad.

\begin{figure}
	\centering
	\includegraphics[width=\linewidth]{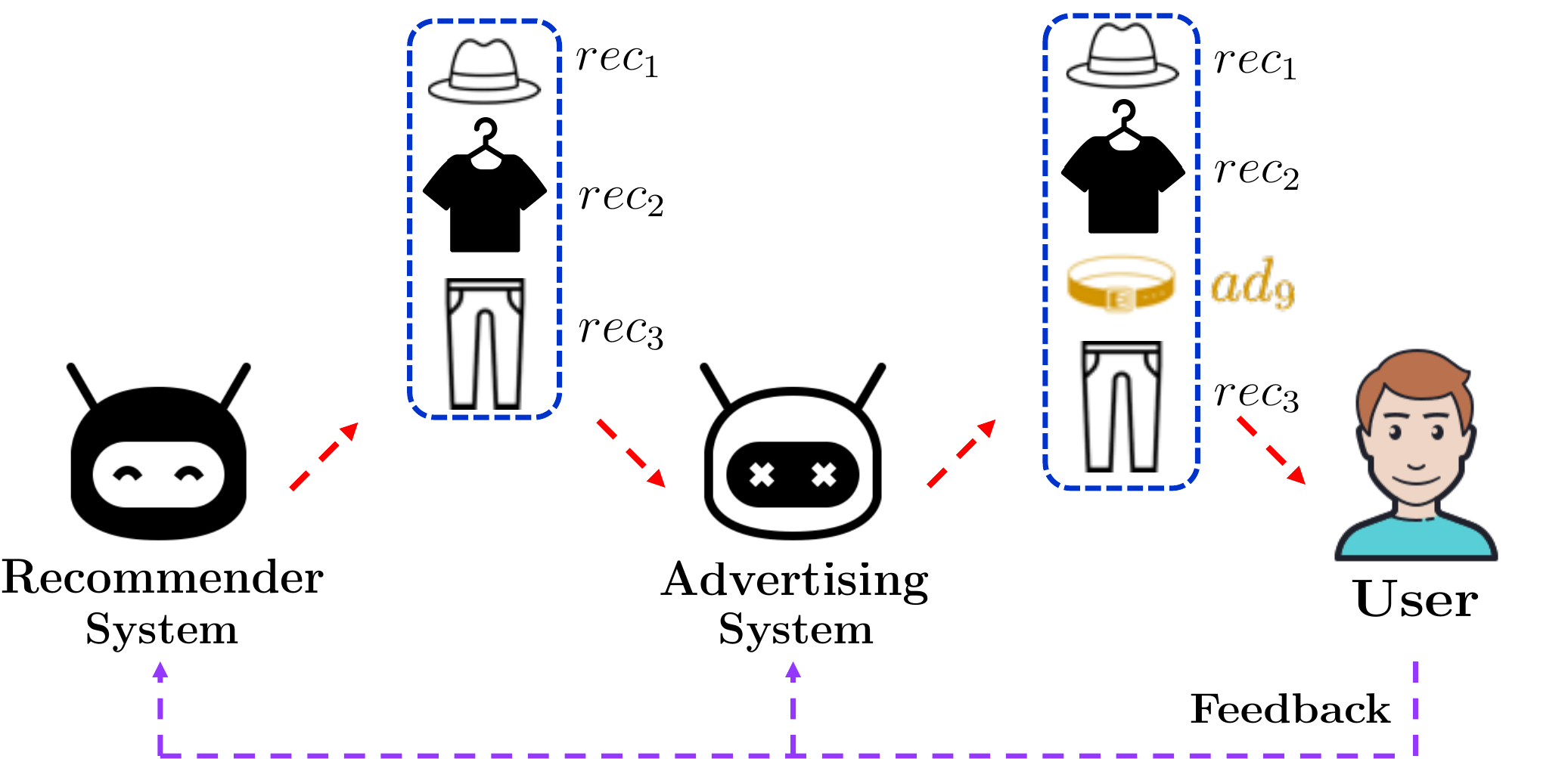}
	\caption{An example of rec-ads mixed display for one user request.}
	\label{fig:Fig1_example}
\end{figure}

To achieve the goal, we propose to study a two-level framework for rec-ads mixed display. Figure~\ref{fig:Fig1_example} illustrates the high-level idea about how the framework works. Upon a user's request, the first level (i.e. RS) firstly generates a list of recommendations (rec-list) according to user's historical behaviors, which aims to optimize the long-term user experience or engagement. The main challenge to build the first level is the high computational complexity of the combinatorial action space, i.e., selecting a subset of items from the large item space. Then the second level (i.e. AS) inserts ads into the rec-list generated from the first level, and it needs to make three decisions, i.e., whether to insert an ad into the given rec-list; and if yes, the AS also needs to decide which ad and where to insert.  For example, in Figure~\ref{fig:Fig1_example}, the AS decides to insert an belt ad $ad_9$ between T-shirt $rec_2$ and pants $rec_3$ of the rec-list. The optimal ad should jointly maximize the immediate advertising revenue from advertisers in the RTB environment and minimize the negative influence of ads on user experience in the long run. Finally, the target user browses the mixed rec-ads list and provides her/his feedback. According to the feedback, the RS and AS update their policies and generate the mixed rec-ads list for the next iteration.

Most existing supervised learning based recommending and advertising methods are designed to maximize the immediate (short-term) reward and suggest items following fixed greedy strategies. They overlook the long-term user experience and revenue. Thus, we build a two-level reinforcement learning framework for \textbf{R}ec/\textbf{A}ds \textbf{M}ixed display (RAM), which can continuously update their recommending and advertising strategies during the interactions with users, and the optimal strategy is designed to maximize the expected long-term cumulative reward from users~\cite{zhao2019deep,zhang2020deep}. Meanwhile, to effectively leverage users' historical behaviors from other policies, we design an off-policy training approach, which can pre-train the framework before launching it online, so as to reduce the bad user experience in the initial online stage when new algorithms have not been well learned~\cite{zhao2019model,zhao2019toward,zou2020neural}. We conduct experiments with real-world data to demonstrate the effectiveness of the proposed RAM framework.
\section{Problem Statement}
\label{sec:problem} 
As aforementioned in Section~\ref{sec:introduction}, we consider the rec/ads mixed display task as a two-level reinforcement learning problem, and model it as a Markov Decision Process (MDP) where the RS and AS sequentially interact with users (environment $\mathcal{E}$) by generating a sequence of rec-ads hybrid-list over time, so as to maximize the cumulative reward from $\mathcal{E}$.  Next,  we define the five elements $(\mathcal{S}, \mathcal{A}, \mathcal{P}, \mathcal{R}, \gamma)$ of  the MDP. 
 
 {\bf State space $\mathcal{S}$}:  A state $s_t \in \mathcal{S}$ includes a user's recommendation and advertisement browsing history before time $t$ and the contextual information of current request at time $t$. The generated rec-list from RS is also considered as a part of the state for the AS. {\bf Action space $\mathcal{A}$}:  $a_t = (a_t^{rs}, a_t^{as}) \in \mathcal{A}$ is the action of RS and AS, where $a_t^{rs}$ of RS is to generate a rec-list, and $a_t^{as}$ of AS is to determine three internally related decisions, i.e., whether to insert an ad in current rec-list;  and if yes, the AS needs to choose a specific ad and insert it into the optimal location of the rec-list. We denote $\mathcal{A}_t^{rs}$ and $\mathcal{A}_t^{as}$ as the rec and ad candidate sets for time $t$, respectively. Without the loss of generality, we assume that the length of any rec-list is fixed and the AS can insert at most one ad into a rec-list. {\bf Reward $\mathcal{R}$}:  After an action $a_t$ is executed at the state $s_t$, a user browses the mixed rec-ads list and provides her feedback. The RS and AS will receive the immediate reward $r_t(s_t,a_t^{rs})$ and $r_t(s_t,a_t^{as})$ based on user's feedback. We will discuss more details about the reward in following sections. {\bf Transition probability $\mathcal{P}$}:  $P(s_{t+1}|s_t,a_t)$ is the state transition probability from $s_t$ to $s_{t+1}$ after executing action $a_t$. The MDP is assumed to  satisfy $P(s_{t+1}|s_t,a_t,...,s_1,a_1) = P(s_{t+1}|s_t,a_t)$. {\bf Discount factor $\gamma$}:  Discount factor $\gamma \in [0,1]$ balances between current and future rewards --  (1) $\gamma = 1$: all future rewards are fully considered into current action; and (2) $\gamma = 0$: only the immediate reward is counted.

\begin{figure}[t]
	\centering
	\includegraphics[width=\linewidth]{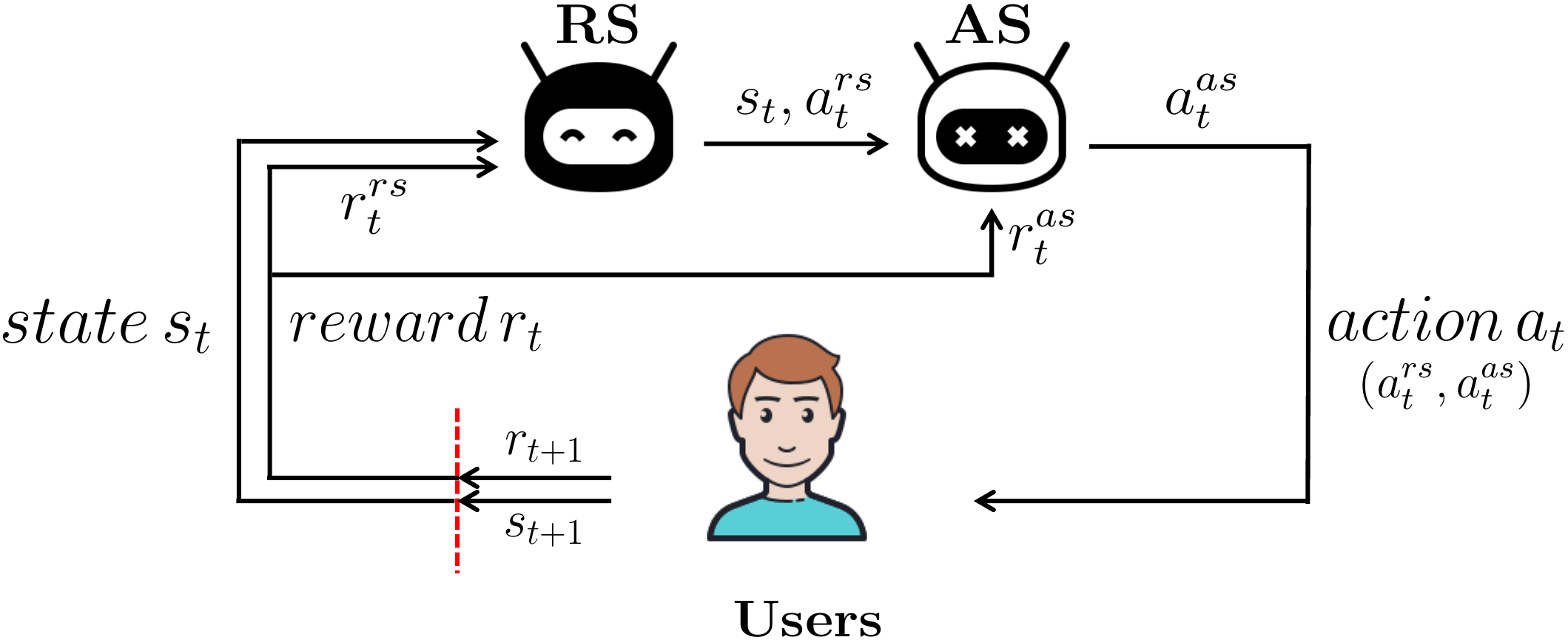}
	\caption{The agent-user interactions in MDP.}
	\label{fig:Fig2_MDP}
\end{figure}

Figure~\ref{fig:Fig2_MDP} illustrates the user-agent interactions in MDP. With the above definitions, the problem can be formally defined as follows: \textit{Given the historical MDP, i.e., $(\mathcal{S}, \mathcal{A}, \mathcal{P}, \mathcal{R}, \gamma)$, the goal is to find a two-level rec/ads policy $\pi = \{\pi_{rs},\pi_{as}\}:\mathcal{S} \rightarrow \mathcal{A}$, which can maximize the cumulative reward from users}, i.e., simultaneously optimizing the user experience and the advertising revenue.
\section{Framework}
\label{sec:framework}
In this section, we will discuss the two-level deep reinforcement learning framework for rec/ads mixed display. We will first introduce the first-level deep Q-network (i.e. RS) to generate a list of recommendations (rec-list) according to user's historical behaviors, then we propose a novel DQN architecture as the second-level (i.e. AS) to insert ads into the rec-list generated from RS. Finally, we discuss how to train the framework via offline data. 

\subsection{Deep Q-network for Recommendations}
\label{sec:architecture_RS}
Given a user request, RS will return a list of items according to user's historical behaviors, which have two major challenges: (i) the high computational complexity of the combinatorial action space $\binom{|\mathcal{A}_t^{rs}|}{k}$, i.e., selecting $k$ items from the large item space $\mathcal{A}_t^{rs}$, and (ii) how to approximate the action-value function (Q-function) for a list of items in the two-level reinforcement learning framework. In this subsection, we introduce and enhance a cascading version of Deep Q-network to tackle the above challenges. Next, we first introduce the processing of state and action features, and then we illustrate the cascading Deep Q-network with an optimization algorithm.

\subsubsection{The Processing of State and Action Features for RS}
\label{sec:features_RS}
As mentioned in Section~\ref{sec:problem}, a state $s_t$ includes a user's rec/ads browsing history, and the contextual information of the current request. The browsing history contains a sequence of recommended items and a sequence of advertised items the user has browsed. Two RNNs with GRU (gated recurrent unit) are utilized to capture users' preferences of recommendations and advertisements, separately.  The final hidden state of RNN is used to represent user's preference of recommended items $p_t^{rec}$ (or ads $p_t^{ad}$). The contextual information $c_t$ of current user request includes app version, operation system (e.g., ios and android) and feed type (swiping up/down the screen), etc.  The state $s_t$ is the concatenation $p_t^{rec}, p_t^{ad}$ and $c_t$ as: 
\begin{equation}\label{equ:s_t}
s_t=concat(p_t^{rec}, p_t^{ad}, c_t)
\end{equation}

For the transition from $s_t$ to $s_{t+1}$, the browsed recommended and advertised items at time $t$ will be inserted into the bottom of $p_{t}^{rs}$ and $p_{t}^{as}$ and we have $p_{t+1}^{rs}$ and $p_{t+1}^{as}$, respectively. For the action $a_t^{rs} = \{a_t^{rs}(1), \cdots, a_t^{rs}(k)\}$ is the embedding of the list of $k$ items that will be displayed in current request. Next, we will detail the cascading Deep Q-network.

\subsubsection{The Cascading DQN for RS}
\label{sec:DQN_RS}
Recommending a list of $k$ items from the large item space $\mathcal{A}_t^{rs}$ is challenging because (i) the combinatorial action space $\binom{|\mathcal{A}_t^{rs}|}{k}$ has high computational complexity, and (ii) the order of items in the list also matters~\cite{zhao2018deep}. For example, a user may have different feedback to the same item if it is placed in different positions of the list. To resolve the above challenges, we leverage a cascading version of DQN which generates a list by sequentially selecting items in a cascading manner~\cite{chen2019generative}. Given state $s_t$, the optimal action is denoted as:
\begin{equation}\label{equ:action1}
\begin{aligned}
a_t^{rs*}  = \{a_t^{rs*}(1), \cdots, a_t^{rs*}(k)\}  = \arg \max _{a_t^{rs}} Q^{*}(s_t, a_t^{rs})
\end{aligned}
\end{equation}

\noindent The key idea of the cascading DQN is inspired by the fact that:
\begin{equation}\label{equ:fact1}
\max _{a_t^{rs}(1:k)} Q^{*}\left(s_t, a_t^{rs}(1{:}k)\right)=\max _{a_t^{rs}(1)}\bigg(\max _{a_t^{rs}(2:k)} Q^{*}\left(s_t, a_t^{rs}(1{:}k)\right)\bigg)
\end{equation}

\noindent which implies a cascade of mutually consistent as:
{\small
\begin{equation}\label{equ:cascading}
\begin{aligned}
a_t^{rs*}(1)&=\arg \max _{a_t^{rs}(1)}\bigg\{Q^{1 *}\left(s_t, a_t^{rs}(1)\right) :=\max _{a_t^{rs}(2:k)} Q^{*}\left(s_t, a_t^{rs}(1{:}k)\right)\bigg\}\\
a_t^{rs*}(2)&=\arg \max _{a_t^{rs}(2)}\bigg\{Q^{2 *}\left(s_t{,}a_t^{rs*}(1){,}a_t^{rs}(2)\right){:=}\max _{a_t^{rs}(3:k)} Q^{*}\left(s_t{,}a_t^{rs}(1{:}k)\right)\bigg\}\\
&\cdots\\
a_t^{rs*}(k)&=\arg\max _{a_t^{rs}(k)}\bigg\{Q^{k *}\left(s_t{,}a_t^{rs*}(1{:}k{-}1){,}a_t^{rs}(k)\right){:=}Q^{*}\left(s_t{,}a_t^{rs}(1{:}k)\right)\bigg\}
\end{aligned}
\end{equation}
}

By applying above functions in a cascading fashion, we can reduce the computational complexity of obtaining the optimal action from $O\binom{|\mathcal{A}_t^{rs}|}{k}$ to $O(k|\mathcal{A}_t^{rs}|)$. Then the RS can sequentially select items following above equations. Note that the items already recommended in the recommendation session will be removed from being recommended again. Next, we will detail how to estimate $\{Q^{j*}|j\in [1,k]\}$.

\subsubsection{The estimation of Cascading Q-functions}
\label{sec:estimation_RS}
\begin{figure}[t]
	\centering
	\includegraphics[width=\linewidth]{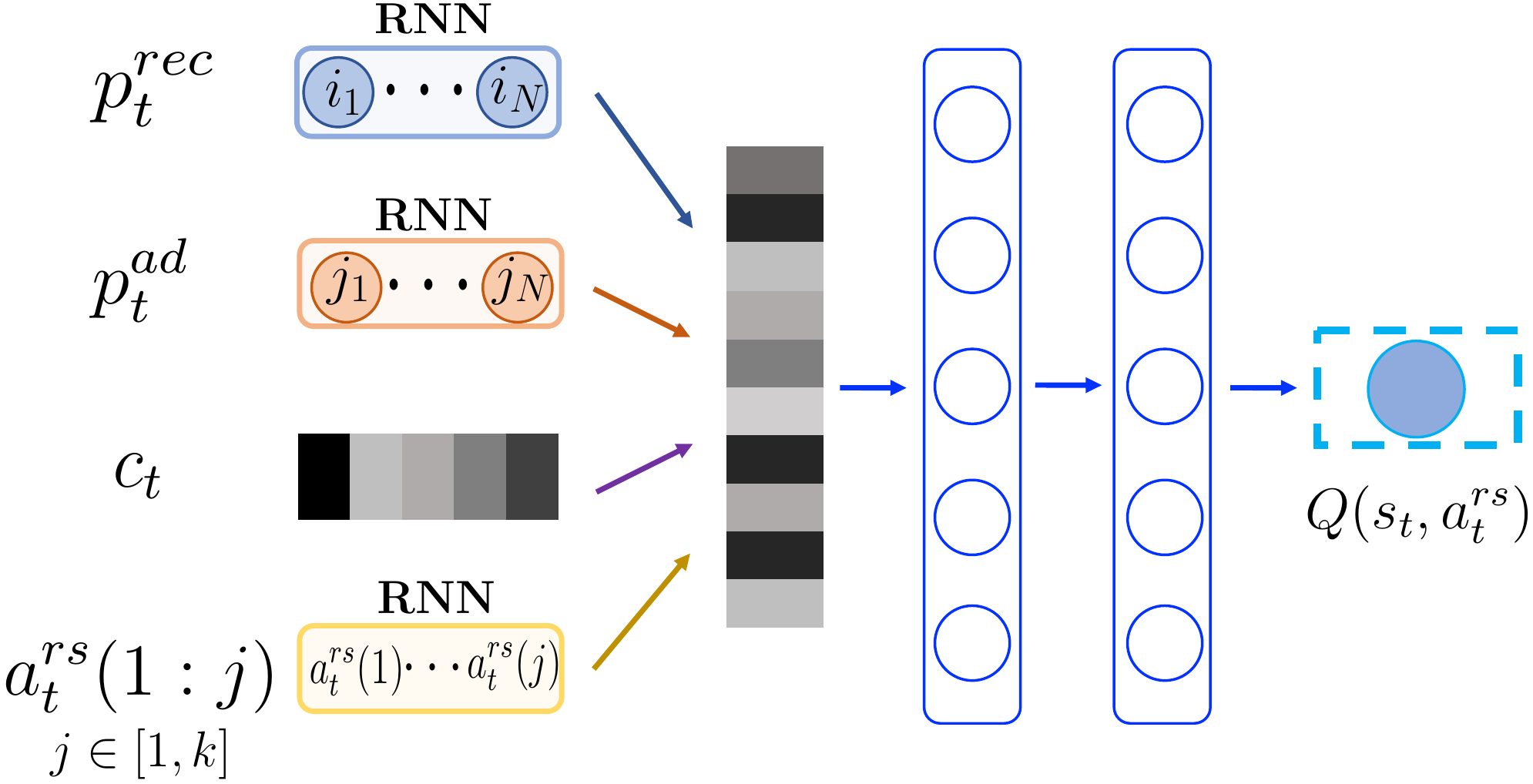}
	\caption{The architecture of cascading DQN for RS.}
	\label{fig:Fig3_DQNRS}
\end{figure}

Figure~\ref{fig:Fig3_DQNRS} illustrates the architecture of the cascading DQN, where $i_1{,}\cdots{,}i_N$ and $j_1{,}\cdots{,}j_N$ are uses' rec and ads browsing histories. The original model in~\cite{chen2019generative} uses $k$ layers to process $k$ items separately without efficient weights sharing, which is crucial in handling large action size~\cite{song2019solving}. To address this challenge, we replace the $k$ separate layers by RNN with GRU, where the input of $j^{th}$ RNN unit is the feature of the $j^{th}$ item in the list, and the final hidden state of RNN is considered as the representation of the list. Since all RNN units share the same parameters, the framework is flexible to any action size $k$.

To ensure that the cascading DQN selects the optimal action, i.e., a sequence of $k$ optimal sub-actions, $\{Q^{j*}|j\in [1,k]\}$ functions should satisfy a set of constraints as follows:
\begin{equation}%\nonumber
Q^{j *}\left(s_t, a_t^{rs*}(1:j)\right)=Q^{*}\left(s_t, a_t^{rs*}(1:k)\right), \quad \forall j \in [1,k]
\end{equation}

\noindent i.e., the optimal value of $Q^{j *}$ should be equivalent to $Q^{*}$ for all $j$. The cascading DQN enforces the above constraints in a soft and approximate way, where the loss functions are defined as follows:
\begin{equation}\label{equ:loss1}
\begin{aligned}
&\left(y_t^{rs}-Q^{j}\left(s_t, a_t^{rs}(1:j)\right)\right)^{2}, \text { where }\\
&y_t^{rs}=r_t\left(s_{t}, a_t^{rs}(1:k)\right)+\gamma Q^{*}\left(s_{t+1}, a_{t+1}^{rs*}(1:k)\right), \forall j \in [1,k]
\end{aligned}
\end{equation}

\noindent i.e., all $Q^{j}$ functions fit against the same target $y_t^{rs}$. Then we update the parameters of the cascading DQN by performing gradient steps over the above loss. We will detail the reward function $r_t\left(s_{t}, a_t^{rs}(1:k)\right)$ in the following subsections. Next, we will introduce the second-level DQN for advertising.

\subsection{Deep Q-network for Online Advertising}
\label{sec:architecture_AS}
As mentioned in Section~\ref{sec:introduction},  the advertising system (AS) is challenging. First, AS needs to make three decisions, i.e., whether, which and where to insert. Second, these three decisions are dependent and traditional DQN architectures cannot be directly applied. For example, only when the AS decides to insert an ad, AS also needs to determine the candidate ads and locations. Third, the AS needs to not only maximize the income of ads but also to minimize the negative influence on user experience. To tackle these challenges, next we detail a novel Deep Q-network architecture. 

\subsubsection{The Processing of State and Action Features for AS}
\label{sec:features_AS}
We leverage the same architecture as that in Section~\ref{sec:features_RS} to obtain the state $s_t$. Furthermore, since the task of AS is to insert ad into a given rec-list, the output of the first-level DQN, i.e., the current rec-list $a_t^{rs} = \{a_t^{rs}(1), \cdots, a_t^{rs}(k)\}$, is also considered as a part of the state for AS. For the action $a_t^{as} = (a_t^{ad}, a_t^{loc})$ of AS, $a_t^{ad}$ is the embedding of a candidate ad.  Given the rec-list of $k$ items, there exist $k+1$ possible locations. Thus, we use a one hot vector $a_t^{loc}\in \mathbb{R}^{k+1}$ to indicate the location to insert the selected ad. 

\begin{figure}[t]
	\centering
	\includegraphics[width=\linewidth]{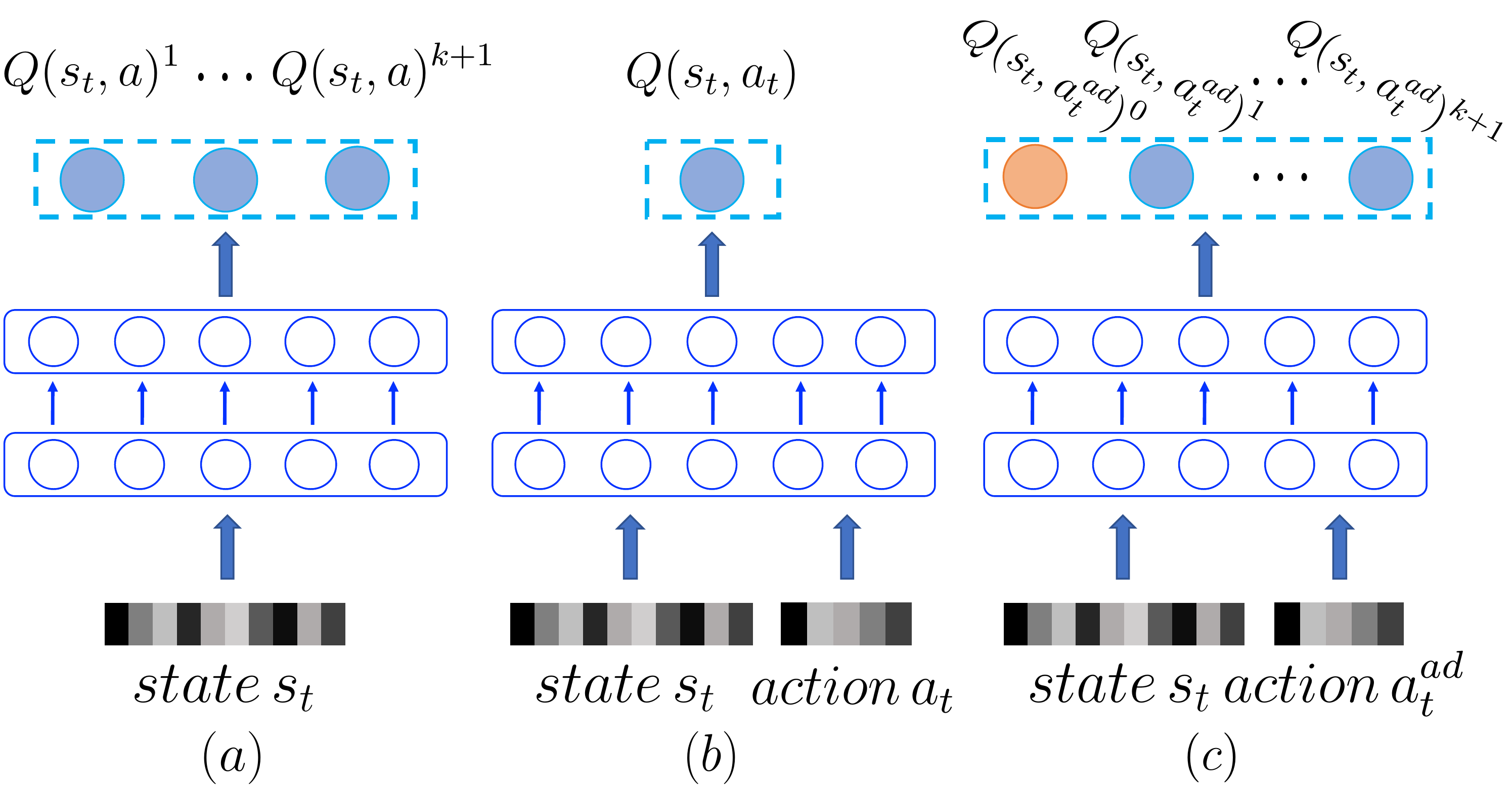}
	\caption{(a)(b) Two conventional DQN architectures. (c) Overview of the proposed DQN Architecture. }
	\label{fig:Fig4_DQNAS}
\end{figure}

\subsubsection{The Proposed DQN Architecture}
\label{sec:DQN}
Given state $s_t$ and rec-list $a_t^{rs}$, the action of AS $a_t^{as}$ contains three sub-actions, i.e., (i) whether to insert an ad into rec-list $a_t^{rs}$; if yes, (ii) which is the best ad and (iii) where is the optimal location. Note that in this work we suppose that the AS can insert an ad into a given rec-list at most. Next we discuss the limitations if we directly apply two classic Deep Q-network (DQN) architectures as shown in Figures~\ref{fig:Fig4_DQNAS}(a) and (b) to the task. The architecture in Figure~\ref{fig:Fig4_DQNAS}(a) inputs only the state ($s_t$ and $a_t^{rs}$) and outputs Q-values of all $k+1$ possible locations. This DQN can select the optimal location, while it cannot choose the optimal ad to insert. The architecture in Figure~\ref{fig:Fig4_DQNAS}(b) takes a pair of state-action and outputs the Q-value for a specific ad. This DQN can determine the optimal ad but cannot determine where to insert the ad. One solution is to input the location information (e.g. one-hot vector). However, it needs $O(k|\mathcal{A}_t^{as}|)$ evaluations (forward propagation) to find the optimal Q-value, which is not practical in real-world advertising systems. Note that neither of the two classic DQNs can decide whether to insert an ad into a given rec-list. 

To address above challenges, we propose a novel DQN architecture for online advertising in a given rec-list $a_t^{rs}$, as shown in Figure~\ref{fig:Fig4_DQNAS}(c).  It is built based on the two classic DQN architectures. The proposed DQN takes state $s_t$ (including $a_t^{rs}$) and a candidate ad $a_t^{ad}$ as input, and outputs the Q-values for all $k+1$ locations. This DQN architecture inherits the advantages of both two classic DQNs. It can evaluate the Q-values of the combination of two internally related types of sub-actions at the same time. In this paper, we evaluate the Q-values of all possible locations for an ad simultaneously. To determine whether to insert an ad (the first sub-action), we extend the output layer from $k+1$ to $k+2$ units, where the $Q(s_t,a_t^{ad})^0$ unit corresponds to the Q-value of not inserting an ad into rec-list $a_t^{rs}$. Therefore, our proposed DQN can simultaneously determine three aforementioned sub-actions according to the Q-value of ad-location combinations $(a_t^{ad}, a_t^{loc})$, and the evaluation times are reduced from $O(k|\mathcal{A}_t^{as}|)$ to $O(|\mathcal{A}_t^{as}|)$; when $Q(s_t,\mathbf{0})^0$ leads to the maximal Q-value, the AS will insert no ad into rec-list $a_t^{rs}$, where we use a zero-vector $\mathbf{0}$ to represent inserting no ad.

\begin{figure}
	\centering
	\includegraphics[width=\linewidth]{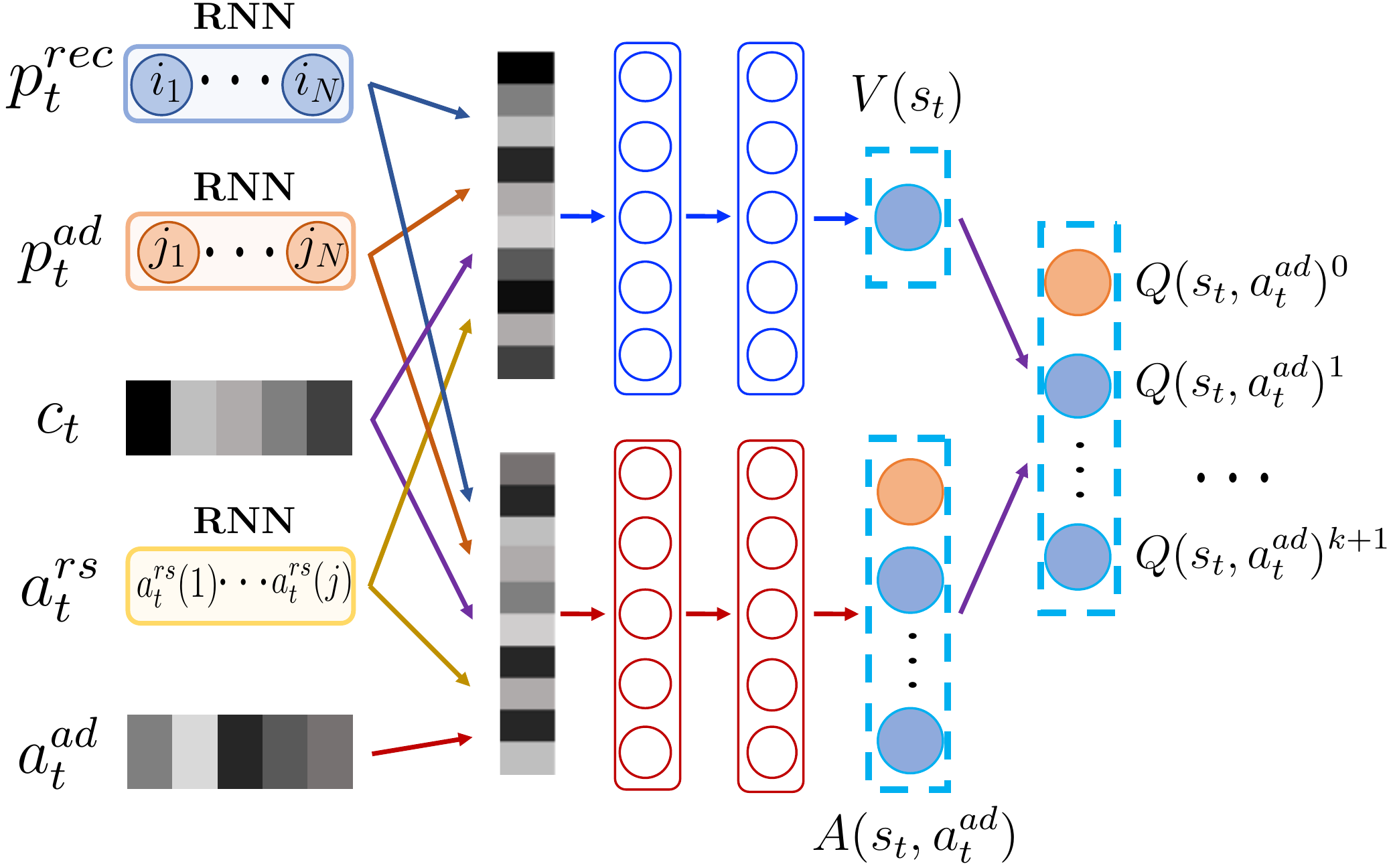}
	\caption{The architecture of the proposed DQN for AS.}
	\label{fig:Fig5_DQNAS1}
\end{figure}

More details of the proposed DQN architecture are illustrated in Figure~\ref{fig:Fig5_DQNAS1}. First, whether to insert an ad into a given rec-list is affected by $s_t$ and $a_t^{rs}$ (especially the quality of the rec-list). For example, if a user is satisfied with a rec-list, the AS may prefer to insert an ad into the rec-list; conversely, if a user is unsatisfied with a rec-list and is likely to leave, then the AS won't insert an ad.  Second, the reward for an ad-location pair is related to all information. Thus, we divide the Q-function into a value function $V(s_t)$, related to $s_t$ and $a_t^{rs}$, and an advantage function $A(s_t, a_t^{ad})$, decided by $s_t$,  $a_t^{rs}$ and $a_t^{ad}$~\cite{Wang2015Dueling}. 

\subsubsection{The Action Selection in RTB setting}
\label{sec:action_selection}
In the real-time bidding environment, each ad slot is bid by advertisers in real-time when an impression is just generated from a consumer visit~\citep{cai2017real}. In other words, given an ad slot, the specific ad to display is determined by the bids from advertisers, i.e. the bidding system (BS), rather than the platform, which aims to maximize the immediate advertising revenue of each ad slot from advertisers. In this paper, as mentioned in Section~\ref{sec:introduction}, the optimal ad selection policy should not only maximize the immediate advertising revenue (controlled by the BS), but also minimize the negative influence of ads on user experience in the long run (controlled by the AS). To achieve this goal, the AS will first calculate the Q-values for all candidate ads and all possible location, referred as to $Q(s_t,\mathcal{A}_t^{as})$, which captures the long-term influence of ads on user experience; and then the BS will select the ad that achieves the trade-off between the immediate advertising revenue and the long-term Q-values:\begin{equation}\label{equ:bs}
a_t^{as} = BS\left(Q(s_t,\mathcal{A}_t^{as})\right)
\end{equation}

\noindent where the operation $Q(s_t,\mathcal{A}_t^{as})$ goes through all candidate ads $\{a_{t}^{ad}\}$ (input layer) and all locations $\{a_{t}^{loc}\}$ (output layer), including the location that represents not inserting an ad. To be more specific, we design two AS+BS approaches as follows:

\begin{itemize}[leftmargin=*]
	\item \textbf{RAM-l}: the optimal ad-location pair $a_t^{as}=(a_t^{ad}, a_t^{loc})$ directly optimizes the linear summation of immediate advertising revenue and long-term user experience:
	\begin{equation}\label{equ:model-1}
	a_t^{as} = \arg\max_{a_t^{as} \in \mathcal{A}_t^{as}} \big(Q(s_t,a_t^{as})+\alpha \cdot rev_t(a_t^{as})\big)
	\end{equation}
	
	where $\alpha$ controls the second term, and $rev_t(a_t^{as})$ is the immediate advertising revenue if inserting an ad, otherwise 0;
	%	\vspace{1mm}
	\item \textbf{RAM-n}: this is a nonlinear approach that the AS first selects a subset of ad-location pairs $\{a_t^{as}\}$ (the size is $N$) that corresponds to optimal long-term user experience $Q(s_t,a_t^{as})$, then the BS chooses one $a_t^{as}$ that has the maximal immediate advertising revenue $rev_t(a_t^{as})$ from the subset. 
\end{itemize}

It is noteworthy that we maximize immediate advertising revenue rather than long-term advertising revenue because: (i) as aforementioned the ad to insert is determined by advertisers rather than the platform (action is not generated by the agent); and (ii) in the generalized-second-price (GSP) setting, the highest bidder pays the price (immediate advertising revenue) bid by the second-highest bidder, if we use immediate advertising revenue as $r_t(s_t, a_t^{as})$, then we cannot select an ad according to its $Q(s_t, a_t^{as})$ that represents the long-term advertising revenue.

\subsection{The Optimization Task}
\label{sec:optimization}
Given a user request and state $s_t$, the RS and AS sequentially generate actions $a_t^{rs}$ and $a_t^{as}$, i.e., a rec-ads hybrid list, and then the target user will browse the list and provide her/his feedback. The two-level framework aims to (i) optimize the long-term user experience or engagement of recommended items (RS), (ii) maximize the immediate advertising revenue from advertisers in RTB environment (BS), and (iii) minimize the negative influence of ads on user long-term experience (AS), where the second goal is automatically achieved by the bidding system, i.e., the advertiser with highest bid price will win the ad slot auction. Therefore, we next design proper reward functions to assist the RL components in the framework to achieve the first and third goals.   

The framework is quite general for the rec-ads mixed display applications in e-commerce, news feed and video platforms. Thus, for different types of platforms, we design different reward functions. For the first level DQN (RS), to evaluate the user experience, we have
\begin{equation}\label{equ:reward1}
	r_t(s_t, a_t^{rs})=
	\left\{\begin{array}{ll}
		 income
		&e{-}commerce\\
		dwell\,time
		& news/videos
	\end{array}\right.
\end{equation}

\noindent where user experience is measured by the income of the recommended items in the hybrid list in e-commerce platforms, and the dwelling time of the recommendations in news/video platforms. Based on the reward function $r_t(s_t, a_t^{rs})$, we can update the parameters of the cascading DQN (RS) by performing gradient steps over the loss in Eq (\ref{equ:loss1}). We introduce separated evaluation and target networks~\citep{mnih2013playing} to help smooth the learning and avoid the divergence of parameters, where $\theta^{rs}$ represents all parameters of the evaluation network, and the parameters of the target network $\theta_T^{rs}$ are fixed when optimizing the loss function. The derivatives of loss function $L(\theta^{rs})$ with respective to parameters $\theta^{rs}$ are presented as follows:
\begin{equation}\label{equ:differentiating1}
\begin{aligned}
\nabla_{\theta^{rs}}L(\theta^{rs}) = &\left(y_t^{rs}{-}Q^{j}\left(s_t{,}a_t^{rs}(1{:}j){;}\theta^{rs}\right)\right)\nabla_{\theta^{rs}}Q^{j}\left(s_t{,} a_t^{rs}(1{:}j){;}\theta^{rs}\right)\\
\end{aligned}
\end{equation}

\noindent where the target $y_t^{rs}=r_t\left(s_{t}{,} a_t^{rs}(1{:}k)\right)+\gamma Q^{*}\left(s_{t+1}{,} a_{t+1}^{rs*}(1{:}k){;}\theta_T^{rs}\right), \\ \forall j \in [1,k]$.

For the second level DQN (AS), since leaving the platforms is the major risk of inserting ads improperly or too frequently, we evaluate user experience by whether user will leave the platform after browsing current rec-ads hybrid list, and we have: 
\begin{equation}\label{equ:reward2}
r_t(s_t, a_t^{as})=
\left\{\begin{array}{ll}
1 & continue\\
0 & leave
\end{array}\right.
\end{equation}

\noindent in other words, the AS will receive a positive reward (e.g. $1$) if the user continues to browse the next list, otherwise $0$ reward. Then the optimal $Q^*(s_t, a_t^{as})$, i.e., the maximum expected return achieved by the optimal policy, follows the Bellman equation~\cite{bellman2013dynamic} as:
\begin{equation}\label{equ:Q*sa}
Q^{*}(s_t, a_t^{as})=r_t(s_t, a_t^{as})+\gamma Q^{*}\big(s_{t+1}, BS\left(Q^{*}(s_{t+1},\mathcal{A}_{t+1}^{as})\right)\big)
\end{equation}

\noindent then the second level DQN can be optimized by minimizing the loss function as:
\begin{equation}\label{equ:loss2}
\begin{aligned}
&\big(y_t^{as}-Q(s_t, a_t^{as})\big)^{2}, \text { where }\\
&y_t^{as}=r_t(s_t, a_t^{as})+\gamma Q^{*}\big(s_{t+1}, BS\big(Q^{*}(s_{t+1},\mathcal{A}_{t+1}^{as})\big)\big)
\end{aligned}
\end{equation}

\noindent where $y_t^{as}$ is the target of the current iteration. We also introduce separated evaluation and target networks~\cite{mnih2013playing} with parameters $\theta^{as}$ and $\theta_{T}^{as}$ for the second level DQN (AS), and $\theta_{T}^{as}$ is fixed when optimizing the loss function in Eq (\ref{equ:loss2}) (i.e. $L(\theta^{as})$). The derivatives of loss function $L(\theta^{as})$ w.r.t. parameters $\theta^{as}$ can be presented as:
\begin{equation}\label{equ:differentiating2}
\nabla_{\theta^{as}}L(\theta^{as}) =\big(y_t^{as}-Q(s_t, a_t^{as};\theta^{as})\big)\nabla_{\theta^{as}}Q(s_t, a_t^{as};\theta^{as})
\end{equation}

\noindent where $y_t^{as}=r_t(s_t, a_t^{as})+\gamma Q^{*}\big(s_{t+1}, BS\big(Q^{*}(s_{t+1},\mathcal{A}_{t+1}^{as};\theta_{T}^{as})\big);\theta_{T}^{as}\big)$. The operation $Q(s_t,\mathcal{A}_t^{as})$ looks through the candidate ad set $\{a_{t+1}^{ad}\}$ and all locations $\{a_{t+1}^{loc}\}$ (including the location of inserting no ad). 

\begin{algorithm}[t]
	\caption{\label{alg:model1} Off-policy Training of the RAM Framework.}
	\raggedright
	{\bf Input}: historical offline logs, replay buffer $\mathcal{D}$\\
	{\bf Output}: well-trained recommending policy $\pi_{rs}^*$ and advertising policy $\pi_{as}^*$\\
	\begin{algorithmic} [1]
		\STATE Initialize the capacity of replay buffer $\mathcal{D}$
		\STATE Randomly initialize action-value functions $Q_{rs}$ and $Q_{as}$
		\FOR{session $=1, M$}
		\STATE  Initialize state $s_{0}$
		\FOR{$t=1, T$}
		\STATE  Observe state $s_t=concat(p_t^{rec}, p_t^{asd}, c_t)$ 
		\STATE  Execute actions $a_{t}^{rs}$ and $a_{t}^{as}$ according to off-policy $b(s_t)$ 
		\STATE  Get rewards $r_t(s_t, a_t^{rs})$ and $r_t(s_t, a_t^{as})$ from offline log
		\STATE  Update the state from $s_t$ to $s_{t+1}$ 
		\STATE  Store $(s_{t}, a_{t}^{rs},a_{t}^{as},r_t(s_t, a_t^{rs}),r_t(s_t, a_t^{as}),s_{t+1})$ transition into the replay buffer $\mathcal{D}$
		\STATE  Sample minibatch of $(s,a^{rs},a^{as},r(s, a^{rs}),r(s, a^{as}),s')$ transitions  from the replay buffer $\mathcal{D}$
		\STATE  Generate RS's next action $a^{rs'}$ according to Eq.(\ref{equ:cascading})
		\STATE  Generate AS's next action $a^{as'}$ according to Eq.(\ref{equ:bs})
		\STATE  $y^{rs}=
		\left\{\begin{array}{ll}
		r(s, a^{rs})
		& \mathrm{terminal\,}  s'\\
		r(s, a^{rs})+\gamma Q_{rs}(s', a^{rs'})
		& \mathrm{non-terminal\,} s'
		\end{array}\right.$
		\STATE  Update parameters $\theta^{rs}$ of $Q_{rs}$ by minimizing  $\big(y^{rs}-Q_{rs}^{j}\left(s, a^{rs}(1:j)\right)\big)^{2}, \forall j \in [1,k]$ via Eq.(\ref{equ:differentiating1})
		\STATE  $y^{as}=
		\left\{\begin{array}{ll}
		r(s, a^{as})
		& \mathrm{terminal\,}  s'\\
		r(s, a^{as})+\gamma Q_{as}(s', a^{as'})
		& \mathrm{non-terminal\,} s'
		\end{array}\right.$
		\STATE  Update parameters $\theta^{as}$ of $Q_{as}$ by minimizing $\big(y^{as}-Q_{as}(s, a^{as})\big)^{2}$ according to Eq.(\ref{equ:differentiating2})
		\ENDFOR
		\ENDFOR
	\end{algorithmic}
\end{algorithm}

\subsection{Off-policy Training}
\label{sec:offline}
Training the two-level reinforcement learning framework requires a large amount of user-system interaction data, which may result in bad user experience in the initial online stage when new algorithms have not been well trained. To address this challenge, we propose an off-policy training approach that effectively utilizes users' historical behaviors log from other policies. The users' offline log records the interaction history between behavior policy $b(s_t)$ (the current recommendation and advertising strategies) and users' feedback. Our RS and AS take the actions based on the off-policy $b(s_t)$ and obtain feedback from the offline log. We present our off-policy training algorithm in detail shown in Algorithm~\ref{alg:model1}.

There are two phases in each iteration of a training session. For the transition generation phase: for the state $s_t$ (line 6), the RS and AS sequentially act $a_{t}^{rs}$ and $a_{t}^{as}$ based on the behavior policy $b(s_t)$ (line 7) according to a standard off-policy way~\cite{degris2012off}; then RS and AS receive the reward $r_t(s_t, a_t^{rs})$ and $r_t(s_t, a_t^{as})$ from the offline log (line 8) and update the state to $s_{t+1}$ (line 9); and finally the RS and AS store transition $(s_{t}, a_{t}^{rs},a_{t}^{as},r_t(s_t, a_t^{rs}),r_t(s_t, a_t^{as}),s_{t+1})$ into the replay buffer $\mathcal{D}$ (line 10). For the model training phase: the proposed framework first samples minibatch of transitions from $\mathcal{D}$ (line 11), then generates actions $a^{rs'}$ and $a^{as'}$ of next iteration according to Eqs.(\ref{equ:cascading}) and~(\ref{equ:bs}) (lines 12-13), and finally updates parameters of $Q_{rs}$ and $Q_{as}$ by minimizing Eqs.(\ref{equ:loss1}) and~.(\ref{equ:loss2}) (lines 14-17). To help avoid the divergence of parameters and smooth the training, we introduce separated evaluation and target Q-networks~\cite{mnih2013playing} . Note that when $b(s_t)$ decides not to insert an ad (line 7), we denote  $a_{t}^{ad}$ as an all-zero vector. 
\begin{algorithm}[t]
	\caption{\label{alg:model2} Online Test of the RAM Framework.}
	\raggedright
	\begin{algorithmic} [1]
		\STATE Initialize action-value functions $Q_{rs}$ and $Q_{as}$ with well-trained weights
		\FOR{session $=1, M$}
		\STATE  Initialize state $s_{0}$
		\FOR{$t=1, T$}
		\STATE  Observe state $s_t=concat(p_t^{rec}, p_t^{asd}, c_t)$ 
		\STATE  Generate action $a_{t}^{rs}$ according to Eq.(\ref{equ:cascading})
		\STATE  Generate action $a_{t}^{as}$ according to Eq.(\ref{equ:bs})
		\STATE  Execute actions $a_{t}^{rs}$ and $a_{t}^{as}$
		\STATE  Observe rewards $r_t(s_t, a_t^{rs})$ and $r_t(s_t, a_t^{as})$ from user
		\STATE  Update the state from $s_t$ to $s_{t+1}$ 
		\ENDFOR
		\ENDFOR
	\end{algorithmic}
\end{algorithm}

\subsection{Online Test}
\label{sec:on_testing}
We present the online test procedure in Algorithm~\ref{alg:model2}. The process is similar to the transition generation stage of Algorithm~\ref{alg:model1}.  Next we detail each iteration of test session as shown in Algorithm~\ref{alg:model2}.  First, the well-trained RS generates a rec-list by $\pi_{rs}^*$ (line 6) according to the current state $s_t$ (line 5). Second, the well-trained AS, collaboratively working with BS, decides to insert an ad into the rec-list (or not) by $\pi_{as}^*$ (line 7).  Third, the reward is observed from the target user to the hybrid list of recommended and advertised items (lines 8 and 9). Finally we transit the state to $s_{t+1}$ (line 10).
\section{Experiment}
\label{sec:experiment}
In this section, we will conduct extensive experiments using data from a short video site to assess the effectiveness of the proposed RAM framework. We first introduce the experimental settings, then compare the RAM framework with state-of-the-art baselines, and finally conduct component and parameter analysis on RAM. 

\subsection{Experimental Settings}
\label{sec:experimental_settings}
Since there are no public datasets consist of both recommended and advertised items, we collected a dataset from a short video site, TikTok, in March 2019. In total, we collect 1,000,000 sessions in chronological order, where the first 70\% is used as training/validation set and the later 30\% is test set. For more statistics of the dataset, please see Table \ref{table:statistics}. There are two types of videos in the dataset: regular videos (recommended items) as well as ad videos (advertised items). The features for a normal video contain: id, like score, finish score, comment score, follow score and group score, where the scores are predicted by the platform. The features for an ad video consist of: id, image size, bid-price, hidden-cost, predicted-ctr and predicted-recall, where the last four are predicted by the platform.  It is worth noting that (i) the effectiveness of the calculated features have been validated in the businesses of the short video site, (ii) we discretize each numeric feature into a one-hot vector, and (iii) baselines are based on the same features for a fair comparison. The implementation code and a dataset sample is available online\footnote{\url{https://www.dropbox.com/sh/6t0n3bmfozx3ypa/AAD5lRgUZQ4FZKG6tBuAz_w1a?dl=0}}.

\begin{table}
	\centering
	\caption{Statistics of the dataset.}
	\label{table:statistics}
	\begin{tabular}{cccc}
		\hline\hline
		session & user & normal video & ad video\\ 
		1,000,000 & 188,409	& 17,820,066 & 10,806,778\\\hline\hline
		{\scriptsize session dwell time} & {\scriptsize session length} & {\scriptsize session ad revenue} & {\scriptsize rec-list with ad}\\
		17.980 min & 55.032 videos& 0.667 & 55.23\%\\\hline\hline
	\end{tabular}
\end{table}

\begin{table*}[]
	\centering
	\caption{Performance comparison.}
	\label{table:result1}
	\begin{tabular}{|c||c||c|c|c|c||c|c|}
		\hline
		\multirow{2}{*}{Metrics} & \multirow{2}{*}{Values} & \multicolumn{6}{c|}{Algorithms} \\ \cline{3-8} 
		&                   & W\&D  & DFM  & GRU  & DRQN  &  RAM-l &  RAM-n \\ \hline\hline
		\multirow{3}{*}{$R^{rs}$} & value & 17.61$\pm$0.16  & 17.95$\pm$0.19  & 18.56$\pm$0.21  & 18.99$\pm$0.18  &  \textbf{19.61$\pm$0.23} &  19.49$\pm$0.16 \\ \cline{2-8} 
		& improv.(\%) & 11.35  & 9.25  & 5.66  & 3.26  & -  & 0.61  \\ \cline{2-8} 
		& p-value & 0.000  & 0.000  & 0.000  & 0.000  & -  & 0.006  \\ \hline\hline
		\multirow{3}{*}{$R^{as}$} & value & 8.79$\pm$0.06  & 8.90$\pm$0.07  & 9.29$\pm$0.09  & 9.37$\pm$0.10 & \textbf{9.76$\pm$0.09}  & 9.68$\pm$0.06  \\ \cline{2-8} 
		& improv.(\%) & 11.03  & 9.66  & 5.06  & 4.16  & -  & 0.83  \\ \cline{2-8} 
		& p-value & 0.000  & 0.000  & 0.000  & 0.000  & -  & 0.009  \\ \hline\hline
		\multirow{3}{*}{$R^{rev}$} & value & 1.07$\pm$0.03  & 1.13$\pm$0.02  & 1.23$\pm$0.04  & 1.34$\pm$0.03  & 1.49$\pm$0.06  &  \textbf{1.56$\pm$0.07} \\ \cline{2-8} 
		& improv.(\%) & 45.81  & 38.05  & 26.83  & 16.42  & 4.70  & -  \\ \cline{2-8} 
		& p-value & 0.000  & 0.000  & 0.000  & 0.000  & 0.001  & -  \\ \hline
	\end{tabular}
\end{table*}

\subsection{Evaluation Metrics}
\label{sec:metrics}
The reward $r_t(s_t, a_t^{rs})$ to evaluate user experience of a list of regular videos is the dwell time (min), and the reward $r_t(s_t, a_t^{as})$ of ad videos is $0$ if users leave the site and $1$ if users continue to browse. We use the \textit{session dwell time} $R^{rs} = \sum_{1}^{T} r_t(s_t, a_t^{rs})$, \textit{session length} $R^{as} = \sum_{1}^{T} r_t(s_t, a_t^{as})$, and session ad revenue $R^{rev} = \sum_{1}^{T} rev_t(a_t^{as})$ as metrics to measure the performance of a test session. 

\subsection{Architecture Details}
\label{sec:architecture}
Next we detail the architecture of RAM to ease reproductivity. The number of candidate regular/ad videos (selected by external recall systems) for a request is $15$ and $5$ respectively, and the size of regular/ad video representation is $60$. There are $k=6$ regular videos in a rec-list. The initial state $s_0$ of a session is collected from its first three requests, and the dimensions of $p_t^{rec}, p_t^{ad}, c_t, a_t^{rs}, a_t^{ad}$ are $64, 64, 13, 360, 60$, respectively. For the second level DQN (AS), two separate 2-layer neural networks are respectively used to generate $V(s_t)$ and $A(s_t, a_t^{ad})$, where the output layer has $k+2=8$ units, i.e., $8$ possible ad locations including not to insert an ad. We empirically set the size of replay buffer 10,000, and the discounted factor of MDP $\gamma = 0.95$. We select important hyper-parameters such as $\alpha$ and $N$ via cross-validation, and we do parameter-tuning for baselines for fair comparison. In the following subsections, we will present more details of parameter sensitivity for the RAM framework. 

\subsection{Overall Performance Comparison}
\label{sec:ev_overall}
The experiment is based on a simulated online environment, which can provide the $r_t(s_t, a_t^{rs})$, $r_t(s_t, a_t^{as})$ and $rev_t(a_t^{as})$ according to a mixed rec-ads list. The simulator shares similar architecture to Figure \ref{fig:Fig5_DQNAS1}, while the output layer predicts the dwell time, whether user will leave and the ad revenue of current mixed rec-ads list. 
We compare the proposed framework with the following representative baseline methods:
\textbf{W\&D} \cite{cheng2016wide}: This baseline jointly trains a wide linear model with feature transformations and a deep feedforward neural networks with embeddings for general recommender systems with sparse inputs. One W\&D estimates the recommending scores of regular videos and each time we recommend $k$ videos with highest scores, while another W\&D predicts whether to insert an ad and estimates the CTR of ads. 
\textbf{DFM} \cite{guo2017deepfm}: DeepFM is a deep model that incorporates W\&D model with factorization-machine (FM). It models high-order feature interactions like W\&D and low-order interactions like FM. 
\textbf{GRU} \cite{hidasi2015session}: GRU4Rec is an RNN with GRU to predict what user will click next according to her/his behavior histories. 
\textbf{DRQN} \cite{hausknecht2015deep}: Deep Recurrent Q-Networks addresses the partial observation problem by considering the previous context with a recurrent structure. DRQN uses an RNN architecture to encode previous observations before the current time. 
Similar to W\&D, we also develop two separate DFMs, GRUs, DRQNs for RS and AS, respectively.

The results are shown in Table~\ref{table:result1}. We make the following observations:
(1) GRU performs better than W\&D and DFM, since W\&D and DFM neglect users' sequential behaviors of one session, while GRU can capture the sequential patterns. 
(2) DRQN outperforms GRU, since DRQN aims to maximize the long-term rewards of a session, while GRU targets at maximizing the immediate reward of each request. This result demonstrates the advantage of introducing RL for online recommendation and advertising.
(3) RAM-l and RAM-n achieve better performance than DRQN, which validates the effectiveness of the proposed two-level DQN framework, where the RS generates a rec-list of recommendations and the AS decides how to insert ads.
(4) RAM-n outperforms RAM-l in session ad revenue, since the second step of RAM-n will select the ad-location pair with maximal immediate advertising revenue, which has a higher probability of inserting ads. 
To sum up, RAM outperforms representative baselines, which demonstrates its effectiveness in online recommendation and advertising.

\begin{figure}[t]
		\centering
		\hspace*{-7mm}
		\includegraphics[width=99mm]{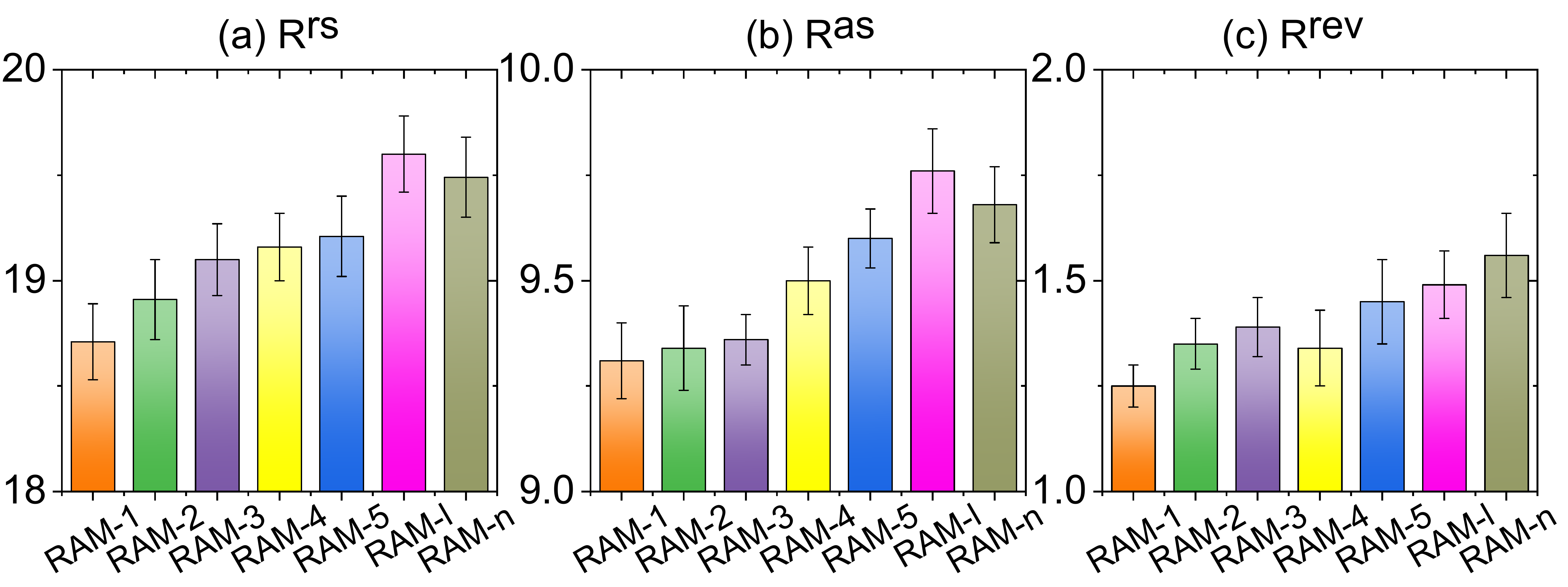}
		\caption{Performance comparison of different variants}
		\label{fig:Fig6_result2}
\end{figure}

\subsection{Component Study}
\label{sec:component}
To understand the impact of model components of RAM, we systematically eliminate the corresponding components of RAM by defining the following variants: 
\textbf{RAM-1}: This variant has the same neural network architectures with the RAM framework, while we train it in the supervised learning way; 
\textbf{RAM-2}: In this variant, we evaluate the contribution of recurrent neural networks, so we replace RNNs by fully-connected layers. Specifically,  we concatenate recommendations or ads into one vector and then feed it into fully-connected layers; 
\textbf{RAM-3}: In this variant, we use the original cascading DQN architecture in~\cite{chen2019generative} as RS; 
\textbf{RAM-4}: For this variant, we do not divide the Q-function of AS into the value function $V(s)$ and the advantage function $A(s, a)$; 
\textbf{RAM-5}: This variant leverages an additional input to represents the location, and uses the DQN in Figure~\ref{fig:Fig4_DQNAS}(b) for AS.

The results are shown in Figure \ref{fig:Fig6_result2}. By comparing RAM and its variants,  we make the following observations: 
(1) RAM-1 demonstrates the advantage of reinforcement learning over supervised learning for jointly optimizing recommendation and online advertising;
(2) RAM-2 validates that capturing user's sequential behaviors can enhance the performance;
(3) RAM-3 proves the effectiveness of RNN over $k$ separate layers for larger action space;
(4) RAM-4 suggests that dividing $Q(s_t, a_t)$ into $V(s_t)$ and $A(s_t, a_t)$ can boost the performance; 
(5) RAM-5 validates the advantage of the proposed AS architecture (over classic DQN architectures) that inputs a candidate ad $a_t^{ad}$ and outputs the Q-value for all possible locations $\{a_t^{loc}\}$.
In summary, leveraging suitable RL policy and proper neural network components can improve the overall performance.

\subsection{Parameter Sensitivity Analysis}
\label{sec:parametric}
Our method has two key hyper-parameters, i.e., (i) the parameter $\alpha$ of RAM-l, and (ii) the parameter $N$ of RAM-n. To study their sensitivities, we fix other parameters, and investigate how the RAM framework performs with the changes of $\alpha$ or $N$.

\begin{figure}[t]
	\centering
	\includegraphics[width=81mm]{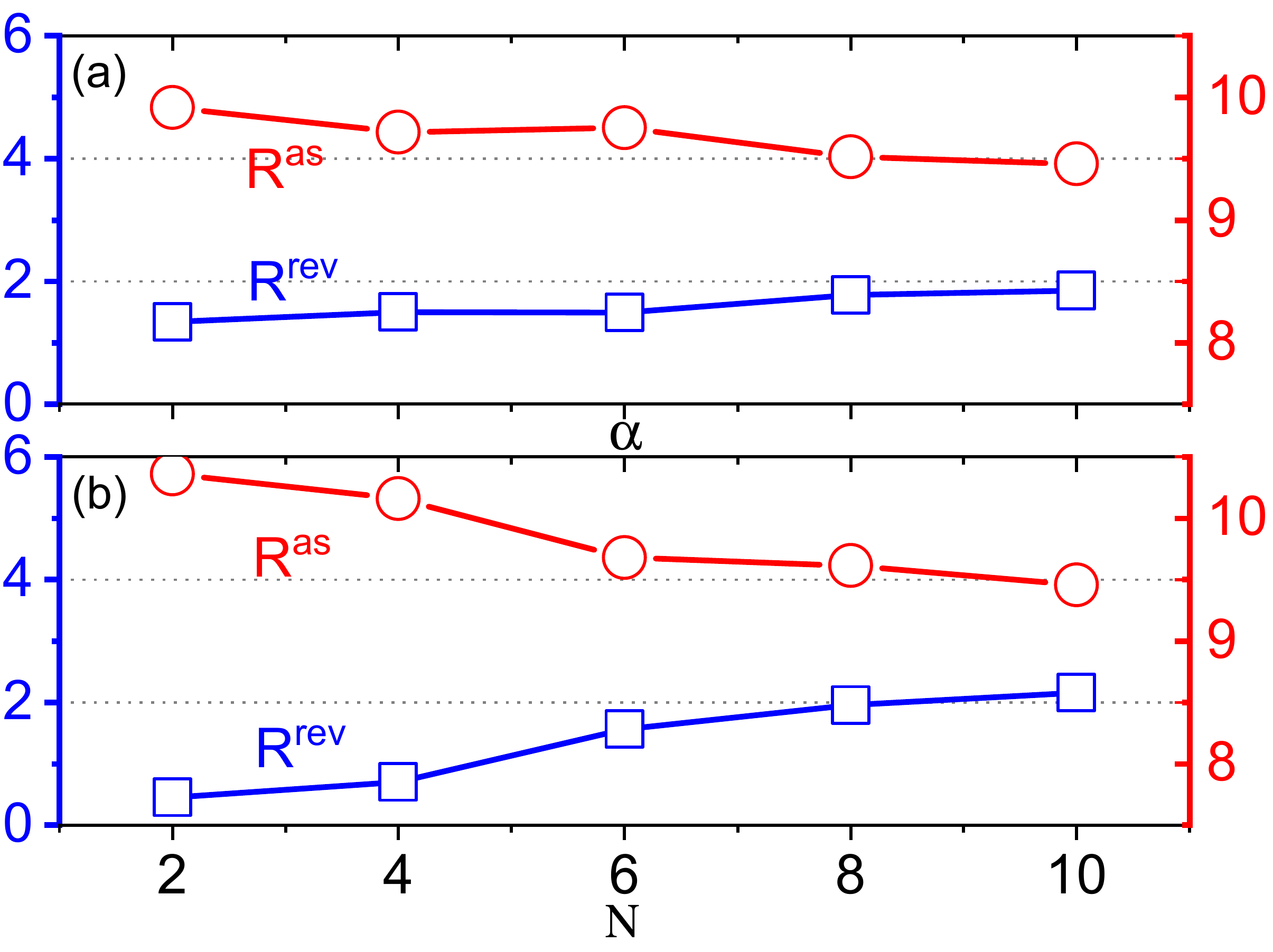}
	\caption{Parameter sensitivity analysis}
	\label{fig:Fig7_result3}
\end{figure}

Figure~\ref{fig:Fig7_result3}(a) illustrates the sensitivity of $\alpha$. We observe that when $\alpha$ increases, the metric $R^{rev}$ improves, while the metric $R^{as}$ decreases. This observation is reasonable because when we decrease the importance of the second term of Equation~(\ref{equ:model-1}), the AS will insert more ads or choose the ads likely to have more revenue, while ignoring their negative impact on regular recommendations.
Figure~\ref{fig:Fig7_result3}(b) shows the sensitivity of $N$. With the increase of $N$, we can observe that the metric $R^{rev}$ improves and the metric $R^{as}$ decreases. With smaller $N$, the first step of RAM-n prefers to selecting most ad-location pairs that not insert an ad, which results in lower $R^{rev}$ and larger $R^{as}$; on the contrary, with larger $N$, the first step returns more pairs with non-zero ad revenue, then the second step leads to higher $R^{rev}$.
In a nutshell, both above results demonstrate that recommended and advertised items are mutually influenced: inserting more ads can lead to more ad revenue while worse user experience, vice versa. Therefore, online platforms should carefully select these hyper-parameters according to their business demands.
\section{Related Work}
\label{sec:related_work}
In this section, we will briefly summarize the related works of our study, which can be mainly grouped into the following categories.

The first category related to this paper is reinforcement learning-based recommender systems. A DDPG algorithm is used to mitigate the large action space problem in real-world RL-based RS~\cite{dulac2015deep}. A tree-structured policy gradient is presented in~\cite{chen2018large} to avoid the inconsistency of DDPG-based RS. Biclustering is also used to model RS as grid-world games to reduce action space~\cite{choi2018reinforcement}. A Double DQN-based approximate regretted reward technique is presented to address the issue of unstable reward distribution in dynamic RS environment~\cite{chen2018stabilizing}. A pairwise RL-based RS framework is proposed to capture users' positive and negative feedback to improve recommendation performance~\cite{zhao2018recommendations}. A page-wise RS is proposed to simultaneously recommend a set of items and display them in a 2-dimensional page~\cite{zhao2018deep,zhao2017deep}. A DQN based framework is proposed to address the issues in the news feed scenario, like only optimizing current reward, not considering labels, and diversity issue~\cite{zheng2018drn}. An RL-based explainable RS is presented to explain recommendations and can flexibly control the explanation quality according to the scenarios~\cite{wang2018reinforcement}. A policy gradient-based RS for YouTube is proposed to address the biases in logged data by introducing a simulated historical policy and a novel top-K off-policy correction~\cite{chen2018top}. 

The second category related to this paper is RL-based online advertising techniques, which belong to two groups. The first group is guaranteed delivery (GD), where ads are charged according to a pay-per-campaign pre-specified number of deliveries~\cite{salomatin2012unified}. A multi-agent RL method is presented to control cooperative policies for the publishers to optimize their targets in a dynamic environment~\cite{wu2018multi}. The second group is real-time bidding (RTB), which allows an advertiser to bid each ad impression in a very short time slot. Ad selection task is typically modeled as multi-armed bandit problem supposing that arms are iid, feedback is immediate and environments are stationary~\cite{nuara2018combinatorial,gasparini2018targeting,tang2013automatic,xu2013estimation,yuan2013adaptive,schwartz2017customer}. The problem of online advertising with budget constraints and variable costs is studied in MAB setting~\cite{ding2013multi}, where pulling the arms of bandit results in random rewards and spends random costs. However, the MAB setting considers the bid decision as a static optimization problem, and the bidding for a given ad campaign would repeatedly happen until the budget runs out. To address these challenges, the MDP setting has also been studied for RTB~\cite{cai2017real,wang2018learning,zhao2018deep,rohde2018recogym,wu2018budget,jin2018real}. A model-based RL framework is proposed to learn bid strategies in RTB setting~\cite{cai2017real}, where state value is approximated by a neural network to better handle the large scale auction volume problem and limited budget. A model-free RL method is also designed to solve the constrained budget bidding problem, where a RewardNet is presented to generate rewards for reward design trap~\cite{wu2018budget}. A multi-agent RL framework is presented to consider other advertisers' bidding as the state, and a clustering method is leveraged to handle the large amount of advertisers issue~\cite{jin2018real}.
\section{Conclusion}
\label{sec:conclusion}
In this paper, we propose a two-level deep reinforcement learning framework RAM with novel Deep Q-network architectures for the mixed display of recommendation and advertisements in online recommender systems. Upon a user's request, the RS (i.e. first level) first recommends a list of items based on user's historical behaviors, then the AS (i.e. second level) inserts ads into the rec-list, which can make three decisions, i.e., whether to insert an ad into the rec-list; and if yes, the AS will select the optimal ad and insert it into the optimal location. The proposed two-level framework aims to simultaneously optimize the long-term user experience and immediate advertising revenue. It is worth to note that the proposed AS architecture can take advantage of two conventional DQN architectures, which can evaluate the Q-value of two kinds of related actions simultaneously. We evaluate our framework with extensive experiments based on data from a short video site TikTok. The results show that our framework can jointly improve recommendation and advertising performance. For future work, the RAM framework is quite general for evaluating the Q-values of two or more types of internally related actions, we would like to investigate its more applications, such as news feed, e-commerce and video games.

\section*{ACKNOWLEDGEMENTS}
This work is supported by National Science Foundation (NSF) under grant numbers IIS1907704, IIS1928278, IIS1714741, IIS1715940, IIS1845081 and CNS1815636.

\bibliographystyle{ACM-Reference-Format}
\bibliography{9Reference} 
\end{document}